\documentclass[useAMS,usenatbib]{mn2e}
\usepackage{epsfig,lscape}

\title[Microvariability properties of BALQSOs]{Optical
  microvariability properties of BALQSOs } \author[R. Joshi et
  al.]{Ravi Joshi$^{1}$\thanks{E-mail: ravi@aries.res.in (RJ);
    hum@aries.res.in (HC); acgupta30@gmail.com (ACG); wiitap@tcnj.edu
    (PJW)}, Hum Chand$^{1}$, Alok C.\ Gupta$^{1}$ and Paul
  J.\ Wiita$^{2}$ \\ $^{1}$Aryabhatta Research Institute of
  Observational Sciences (ARIES), Manora Peak, Nainital, 263129,
  India\\ $^{2}$Department of Physics, The College of New Jersey, PO
  Box 7718, Ewing, NJ 08628, USA}
\begin{document}
\date{Accepted 2010 November 25. Received 2010 November 02; in original form 2010 September 07}

\pagerange{\pageref{firstpage}--\pageref{lastpage}} \pubyear{2009}

\maketitle

\label{firstpage}
\begin{abstract}
We present optical light curves of 19 radio quiet (RQ) broad
absorption line (BAL) QSOs and study their rapid variability
characteristics. Systematic CCD observations, aided by a careful
data analysis procedure, have allowed us to clearly detect any such
microvariability exceeding 0.01--0.02 mag.  Our observations cover a
total of 13 nights ($\sim$72 hours) with each quasar monitored for
about 4 hours on a given night.  Our sample size is a factor of
three larger than the number of radio-quiet BALQSOs previously
searched for microvariability.  We introduce a scaled $F-$test
statistic for evaluating the presence of optical microvariability
and demonstrate why it is generally preferable to the statistics
usually employed for this purpose.  Considering only unambiguous
detections of microvariability we find that $\sim$11 per cent of
radio-quiet BALQSOs (two out of 19 sources) show microvariability
for an individual observation length of about 4 hr.  This new duty
cycle of 11\% is similar to the usual low microvariability fraction
of normal RQQSOs with observation lengths similar to those of
ours. This result provides support for models where radio-quiet
BALQSO do not appear to be a special case of the RQQSOs in terms of
their microvariability properties.
\end{abstract}
\section{Introduction}
Significant variability in brightness over a few minutes to several
hours (less than a day) is commonly known as microvariability,
intra-night optical variability (INOV) or intra-day
variability. Optical microvariability is a well known property of
radio-loud (RL) active galactic nuclei (AGN), particularly of its
blazar subclass (e.g., Gupta et al.\ 2008 and references
therein). Over past two decades there have been rather extensive
searches for this phenomenon in blazars, other types of RLQSOs, and
the far more numerous radio quiet quasars (RQQSOs) (e.g., Miller,
Carini \& Goodrich 1989; Carini et al.\ 1992, 2007; Gopal-Krishna et
al.\ 1993b, 2000, 2003; de Diego et al.\ 1998; Romero, Cellone \&
Combi 1999; Sagar et al.\ 2004; Stalin et al.\ 2004; Montagni et
al.\ 2006; Goyal et al.\ 2010).  In the case of blazars these studies
have provided useful constraints on the relativistic jet based models
that are used to explain the origin of the large variations that help
define the category (e.g., Marscher, Gear \& Travis 1992; Rani et
al.\ 2010).  Since RQQSOs lack jets of significant power and extent,
the microvariability seen in them may arise from processes on the
accretion disc itself, and thus could possibly be used to probe the
properties of the discs (e.g., Gopal-Krishna, Sagar \& Wiita 1993a).
However, so far there has been a lack of systematic effort to exploit
microvariability properties to understand the nature of the
substantial quasar sub-class with broad absorption lines (BALs), the
BALQSOs. \par

These BALQSOs are AGN characterized by the presence of strong
absorption troughs in their optical spectra. They constitute about
10--15 per cent of optically selected quasars (e.g., Reichard et
al.\ 2003; Hewett \& Foltz 2003). The BALs are attributed to material
flowing outwards from the nucleus with velocities of 5000 to 50000 km
s$^{-1}$ (Green et al.\ 2001). BALQSOs are classified mainly into two
subclasses based on the material predominantly producing the BAL
troughs. High ionization BAL quasars (HiBALs) have broad absorption
from C IV, Si IV, N V and O IV lines. About 10 per cent of BALQSOs
also show, along with HiBAL features, broad absorptions from lower
ionization lines such as Mg II or Al III; these are called
low-ionization BAL quasars (LoBALs).  Any complete model of quasars
and AGNs needs to explain self-consistently a wide range of their
observational properties which also include: the presence of other
emission/absorption lines, the fraction of quasars showing broad
absorption lines and the fraction among them showing continuum
variability such as microvariability.

Carini et al.\ (2007) have compiled a sample of 117 radio-quiet
objects that have been searched for their microvariability.  Of these,
47 are classified as Seyfert galaxies, 64 as QSOs, and 6 as BALQSOs.
In their entire sample 21.4 per cent of the objects were found to
exhibit microvariability, but among objects classified as Seyfert
galaxies, QSOs and BALQSOs, microvariability was seen in 17 per cent,
23 per cent and 50 per cent, respectively (Carini et al.\ 2007).  In
addition, Rabbette et al.\ (1998) have noted that two radio-quiet
BALQSOs displayed short term X-ray variability.  The observed high
fraction of microvariations in BALQSOs suggests that it might be
worthwhile to expend more of the observing time devoted to
microvariability on the BALQSO class if one wants to understand
physical processes in or near the accretion disc.  Clearly, the
present sample size of BALQSOs is very small compared to those of the
non-BALQSO classes, and no useful conclusions about their nature can
be drawn from them. Therefore it is important to increase the sample
of BALQSOs, so as to be able to arrive at firmer conclusions about the
fraction showing microvariability.  We note that if BALQSOs do really
show a substantially higher duty cycle for microvariability than do
non-BAL RQQSOs, this would shed light on the question of whether or
not radio-quiet BALQSOs are special cases of the RQQSOs, especially in
terms of their microvariability properties. For instance, if even weak
jets dominate the rapid variability (e.g., Gopal-Krishna et al.\ 2003)
then a higher duty cycle for microvariability will give indirect
support for the hypothesis that BALQSOs are viewed at angles nearly
perpendicular to their accretion discs (e.g., Ghosh \& Punsly 2007).
This is because jet fluctuations originating in relativistic jets
pointing close to our line-of-sight are amplified in magnitude and
compressed in timescale (e.g., Gopal-Krishna et al.\ 2003).  Whereas,
if BALQSOs show only the usual low microvariability duty cycles of
normal RQQSOs (around 20 per cent) and the fluctuations still arise in
weak jets, that would provide indirect support for alternative models,
such as those where the BAL outflows come out closer to the disc plane
(e.g., Elvis 2000).  In conjunction with X-ray and optical spectral
properties, such variation information is very useful in constraining
various physical models for the origin of microvariability (e.g.,
Czerny et al.\ 2008) and the nature of BALQSOs itself (e.g., Weymann
et al.\ 1991; Elvis et al. 2000). To address these questions, we have
recently started a pilot program to make an extensive search for
optical microvariability of BALQSOs.  \par

This paper is organized as follows. In Section 2 we describe the main
aspects of our sample selection criteria, while Section 3 briefly
describes our observations and the data reductions. In Sections 4 and
5 we present our analysis and results, respectively. Section 6 gives
a discussion and our conclusions.

\begin{table*}
\centering
\begin{minipage}{180mm}
\caption{Properties of the observed BALQSOs.}
\label{tab:sample}
\begin{tabular}{@{}llccccccrccl@{}}
\multicolumn{1}{c}{Object{\footnote{Sources marked with $\dagger$ were
      observed from the 2.01-m Himalayan Chandra Telescope (HCT), the
      others from the ARIES 1.04-m telescope.  \\ $\star$ Apparent
      $R_{mag}$ and absolute magnitude are taken from Veron-Cetty \&
      Veron (2006).}}}  &\multicolumn{1}{c}{$\alpha_{2000.0}$}
&\multicolumn{1}{c}{$\delta_{2000.0}$} &{g$_{i}$} &{M$_{i}$}
&{$z_{em}$} &{R\footnote{Ratio of the radio [5 GHz] flux to the
    optical [2500\AA] flux taken from SDSS DR7 (Schneider et
    al.\ 2010); ND means no radio detection.}}  &{Type\footnote{BAL
    type: for HiBAL, LoBAL see text; MiBAL = Mini Broad Absorption
    Line Quasar.}}  &{Ref \footnote{References: (1) Gibson et
    al.\ (2009); (2) Scaringi et al.\ (2009); (3) Weymann et
    al.\ (1991); (4) Trump et al.\ (2006) }} &{Date of Obs} \\ \hline
\hline WFM91 0226$-$1024 &02$^h$ 28$^m$ 39.20$^s$ &$-$10$^h$ 11$^m$
10.0$^s$ &15.16$^{\star}$ &$-$30.7$^{\star}$ &2.256& 0.38 &HiBAL & 3 &
22.12.2009 \\ J073739.96$+$384413.2 &07$^h$ 37$^m$ 39.96$^s$
&$+$38$^h$ 44$^m$ 13.2$^s$ &16.99 &$-$28.04 &1.399& ND &LoBAL & 1 &
22.12.2009 \\ J084044.41$+$363327.8 &08$^h$ 40$^m$ 44.41$^s$
&$+$36$^h$ 33$^m$ 27.8$^s$ &16.59 &$-$28.36 &1.225& 0.66 &LoBAL & 1 &
06.01.2010 \\ J084538.66$+$342043.6 &08$^h$ 45$^m$ 38.66$^s$
&$+$34$^h$ 20$^m$ 43.6$^s$ &16.96 &$-$29.03 &2.149& ND &HiBAL & 1 &
07.01.2010 \\ J090924.01$+$000211.0 &09$^h$ 09$^m$ 24.01$^s$
&$+$00$^h$ 02$^m$ 11.0$^s$ &16.68 &$-$29.12 &1.864& ND &HiBAL & 4 &
25.01.2010 \\ J094443.13$+$062507.4 &09$^h$ 44$^m$ 43.13$^s$
&$+$06$^h$ 25$^m$ 07.4$^s$ &16.25 &$-$27.40 &0.695& 0.14 &LoBAL & 1 &
16.02.2010 \\ J094941.10$+$295519.2 &09$^h$ 49$^m$ 41.10$^s$
&$+$29$^h$ 55$^m$ 19.0$^s$ &16.04 &$-$28.56 &1.665& ND &HiBAL & 2 &
07.01.2010 \\ J100711.81$+$053208.9 &10$^h$ 07$^m$ 11.81$^s$
&$+$05$^h$ 32$^m$ 08.9$^s$ &16.21 &$-$29.71 &2.143& ND &HiBAL & 1 &
25.03.2010 \\ J111816.95$+$074558.1 &11$^h$ 18$^m$ 16.95$^s$
&$+$07$^h$ 45$^m$ 58.1$^s$ &16.27 &$-$29.34 &1.735& ND &MiBAL & 4 &
25.01.2010 \\ J112320.73$+$013747.4 &11$^h$ 23$^m$ 20.70$^s$
&$+$01$^h$ 37$^m$ 47.0$^s$ &15.84 &$-$29.32 &2.130& ND &LoBAL & 2 &
17.01.2010 \\ J120051.52$+$350831.6 &12$^h$ 00$^m$ 51.52$^s$
&$+$35$^h$ 08$^m$ 31.6$^s$ &16.79 &$-$28.77 &1.717& 0.23 &HiBAL & 1 &
12.05.2010 \\ J120924.07$+$103612.0 &12$^h$ 09$^m$ 24.07$^s$
&$+$10$^h$ 36$^m$ 12.0$^s$ &16.53 &$-$25.69 &0.394& 0.33 &LoBAL & 1 &
08.05.2010 \\ J123820.19$+$175039.1 &12$^h$ 38$^m$ 20.19$^s$
&$+$17$^h$ 50$^m$ 39.1$^s$ &16.86 &$-$25.99 &0.449& 1.03 &LoBAL & 2 &
09.05.2010 \\ J125659.92$+$042734.3 &12$^h$ 56$^m$ 59.90$^s$
&$+$04$^h$ 27$^m$ 34.0$^s$ &15.80 &$-$28.19 &1.025& ND &LoBAL & 2 &
16.02.2010 \\ J151113.84$+$490557.4$\dagger$&15$^h$ 11$^m$ 13.84$^s$
&$+$49$^h$ 05$^m$ 57.4$^s$ &16.49 &$-$28.37 &1.359& 0.91 &LoBAL & 1 &
23.04.2010 \\ J152350.42$+$391405.2$\dagger$&15$^h$ 23$^m$ 50.42$^s$
&$+$39$^h$ 14$^m$ 05.2$^s$ &16.68 &$-$26.77 &0.661& 1.01 &LoBAL & 1 &
24.04.2010 \\ J152553.89$+$513649.1 &15$^h$ 25$^m$ 53.89$^s$
&$+$51$^h$ 36$^m$ 49.1$^s$ &16.85 &$-$30.03 &2.882& ND &HiBAL & 1 &
12.05.2010 \\ J154359.44$+$535903.2 &15$^h$ 43$^m$ 59.44$^s$
&$+$53$^h$ 59$^m$ 03.2$^s$ &17.03 &$-$29.28 &2.370& ND &HiBAL & 1 &
09.05.2010 \\ J160207.68$+$380743.0 &16$^h$ 02$^m$ 07.70$^s$
&$+$38$^h$ 07$^m$ 43.1$^s$ &16.10 &$-$28.62 &1.594& ND &LoBAL & 1 &
14.06.2010 \\ \\ \hline
\end{tabular}
\end{minipage}
\end{table*}  

\section{Source selection criteria}
Our sample is chosen from the BALQSO catalogues compiled by Trump et
al.\ (2006), Scaringi et al.\ (2009) and Gibson et al.\ (2009) which
are based on Sloan Digital Sky Survey (SDSS) Data Releases 3 and 5
(DR3: Schneider et al.\ 2005; DR5: Adelman-McCarthy et al.\ 2007;
Schneider et al.\ 2007). In addition, we also included one brighter
BALQSO from the compilation by Weymann et al.\ (1991). Most of the
sources were selected in such a way that both optical and X-ray
spectral data are available for them in archives.  All were at
declinations that allowed for the observations to be made at
relatively low air masses.  We also required our candidate sources to
have g$_{i} \leq 17$.  This constraint means that even with a 1-m
class telescope we could obtain a good enough signal to noise ratio to
detect fluctuations of $<0.02$ mag with a reasonably good time
resolution of $< 10$ minutes. We also limit the BALQSOs to have
absolute magnitudes M$_{i} < -24.5$, so that the flux contribution
from the host galaxy can be assumed to be negligible (Miller et
al.\ 1990).  \par Our final sample consists of a total of 19 BALQSOs,
as listed in Table~\ref{tab:sample}. Among these BALQSOs 8 are
classified as HiBALs, 10 are LoBALs and 1 is a mini BAL. The whole
sample covers a redshift range of $ 0.39< z_{em} < 2.9$.
  
%%%%%%%%%%%%%%%%%%%%%FIGURE%%%%%%%%%%%%%%%%%%%%%%%%%%%%%%%%%%%%%%

\begin{figure*}
\epsfig{figure=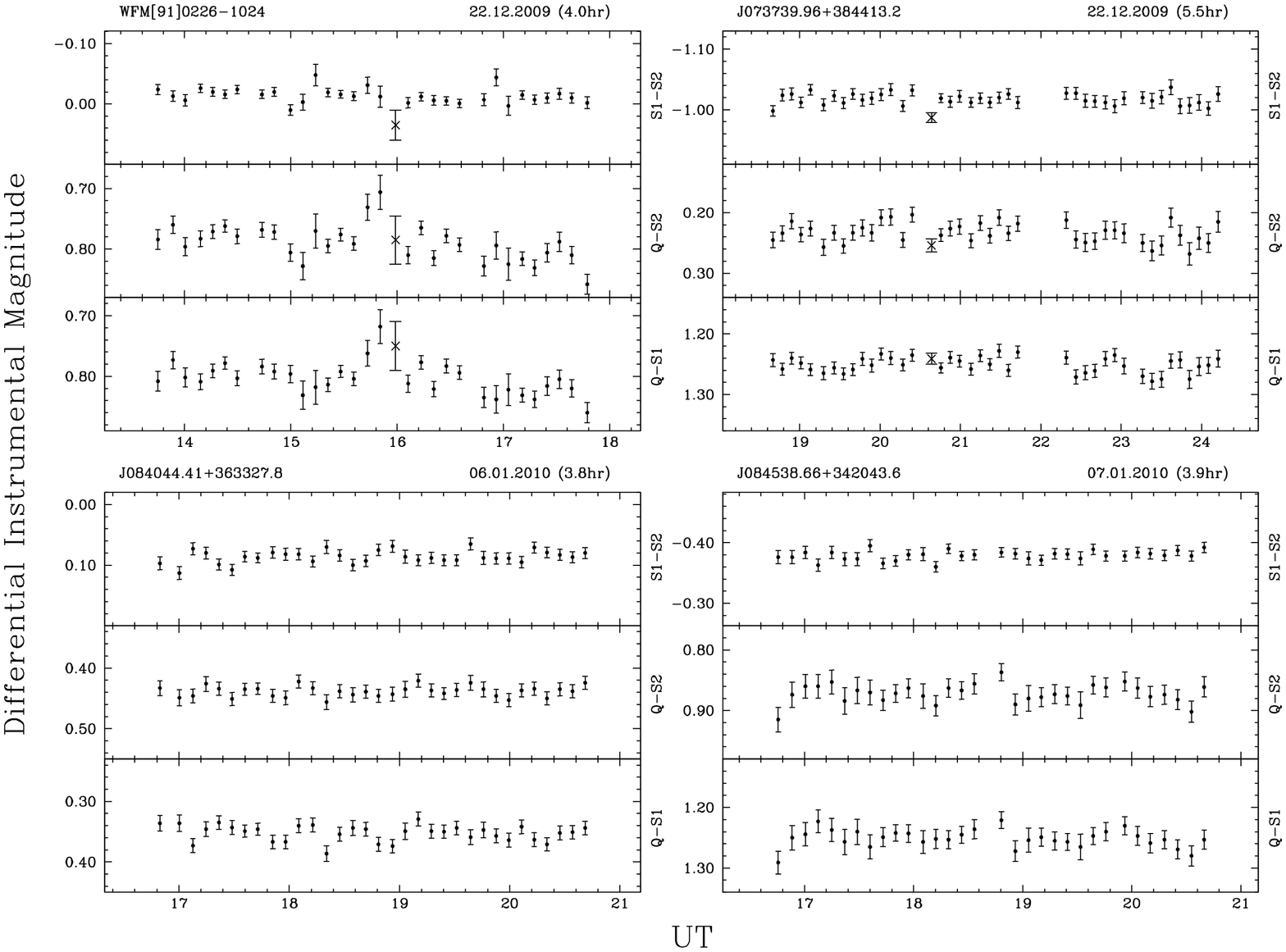,height=18.cm,width=12.cm,angle=270}
 \caption{Differential light curves (DLCs) for the first four BALQSOs
   in our sample.  The name of the quasar and the date and duration of
   the observation are given at the top of each night's data. The
   upper panel gives the comparison star-star DLC and the subsequent
   lower panels give the quasar-star DLCs, as defined in the labels on
   the right side. Any likely outliers (at $> 3\sigma$) in the
   star-star DLCs are marked with crosses, and those data are not used
   in our final analysis.}
\label{fig:s01to04}
\end{figure*}

\begin{figure*}
\epsfig{figure=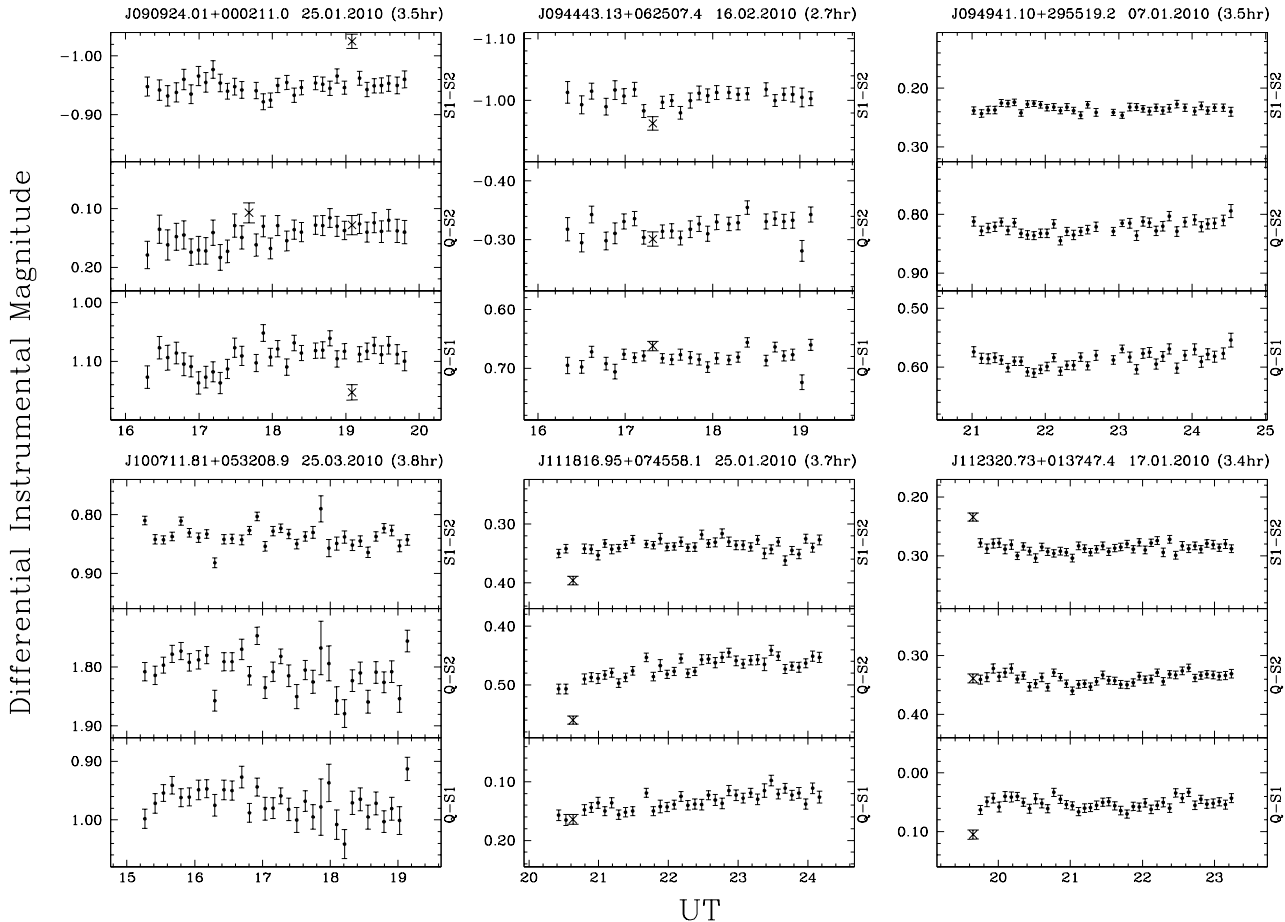,height=18.cm,width=13.cm,angle=270} 
 \caption{As in Fig.~\ref{fig:s01to04} for 6 more BALQSOs.}
\label{fig:s05to10}
\end{figure*}

\begin{figure*}
  \epsfig{figure=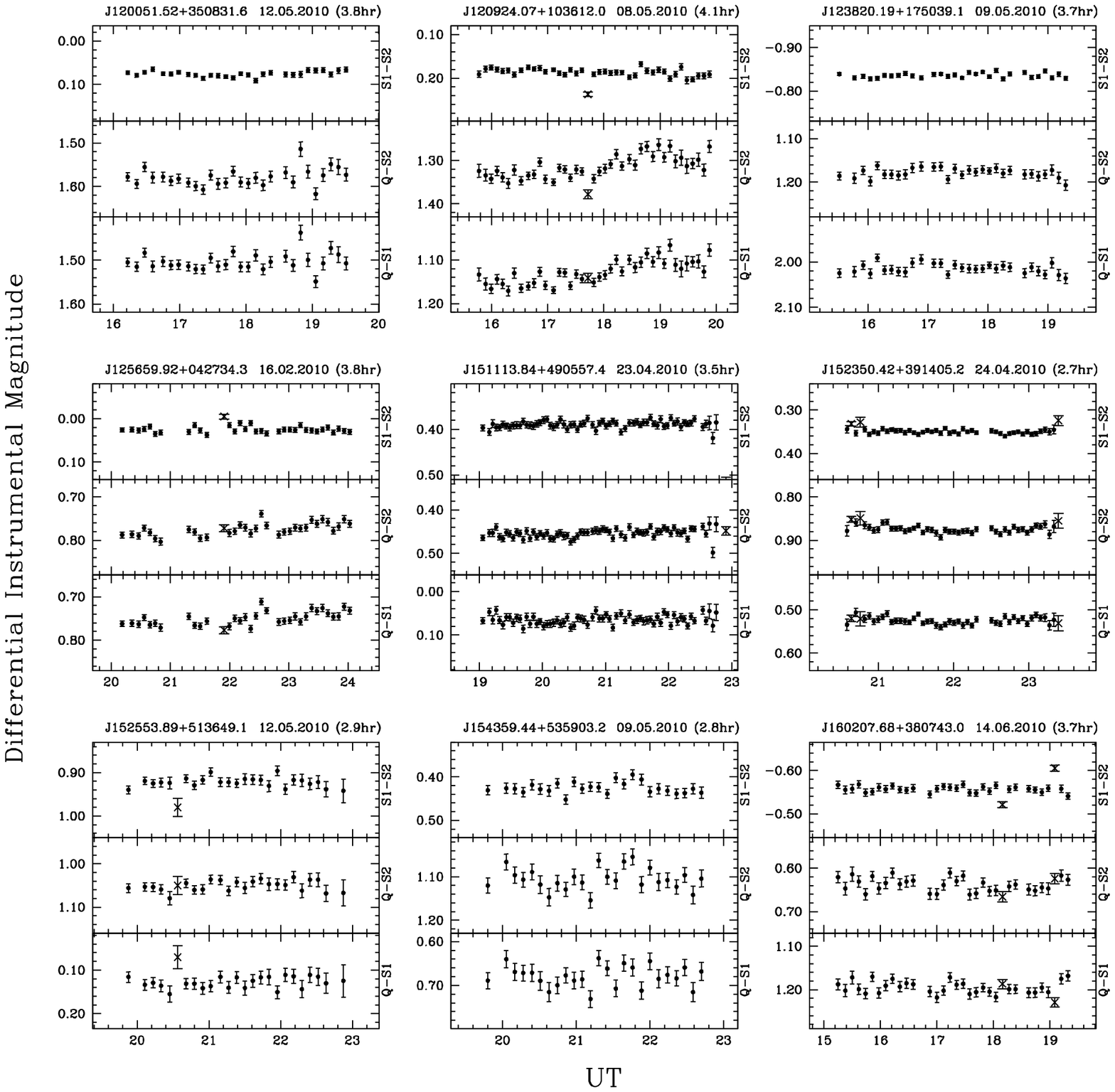,height=18.cm,width=18.cm,angle=0} 
 \caption{As in Fig.~\ref{fig:s01to04} for our last 9 BALQSOs.}
\label{fig:s11to19}
\end{figure*}

%%%%%%%%%%%%%%%%%%%%%FIGURE%%%%%%%%%%%%%%%%%%%%%%%%%%%%%%%%%%%%%%

\section{Observations and Data Reductions}
\subsection{Photometric observations}
Our observations of each of the BALQSOs were carried out continuously
for $\sim$ 4h in the R passband, mainly using the 1.04-m Sampurnanand
telescope located at the Aryabhatta Research Institute of
observational sciencES (ARIES), Nainital, India. It has
Ritchey-Chretien (RC) optics with a f$/$13 beam and is equipped with a
cryogenically cooled CCD detector with a 2048 pixel $\times$ 2048
pixel chip mounted at the Cassegrain focus (Sagar 1999). The readout
noise of the CCD chip is 5.3 e$^{-}$/pixel and it has a gain of 10
e$^{-}$$/$ Analog to Digital Unit (ADU).  Each pixel of the CCD chip
has a dimension of 24 $\mu$m$^{2}$, corresponding to 0.37 arcsec$^{2}$
on the sky, and so covers a total field of $\sim$ 13$^{\prime}$
$\times$ 13$^{\prime}$. To improve the signal to noise ratio,
observations were carried out in a 2 pixel $\times$ 2 pixel binning
mode. The typical seeing during our observing runs at ARIES was $\sim$
3$^{\prime\prime}$.  In addition, two sources were observed with
2.01-m Himalayan Chandra Telescope (HCT) located at the Indian
Astronomical Observatory (IAO), Hanle, India. It is also of the RC
design with a f$/$9 beam at the Cassegrain
focus\footnote{http://www.iiap.res.in/$\sim$iao}.  The detector was a
cryogenically cooled 2048 $\times$ 4096 chip, of which the central
2048 $\times$ 2048 pixels were used. The pixel size is 15 $\mu$m$^{2}$
so that the image scale of 0.29 arcsec$/$pixel covers an area of about
10$^{\prime}$ $\times$ 10${^\prime}$ on the sky. The readout noise of
this CCD is 4.87 e$^{-}$/pixel and the gain is 1.22 e$^{-}$$/$ADU. The
CCD was used in an unbinned mode. The typical seeing during our
observations at IAO was $\sim$ 1.5$^{\prime\prime}$. \par

We chose an R filter for this observational program because it is at
the maximum response of the CCD system; thus the time resolution
achievable for each object is maximized. As most of our sources have
$g_{i}\sim 16-17$, the best time resolution we could achieve was of
the order of 3 minutes, and we almost always managed data points
spaced less than 8 minutes apart, so very rapid fluctuations could be
picked up.  We also took care to select sources and fields of view so
as to ensure availability of at least two, but usually more,
comparison stars on the CCD frame that were within around 1 mag of the
QSO's brightness. This allowed us to identify and discount any
comparison star which itself varied during a given night and hence
ensured reliable differential photometry of the QSO. Observations were
made on a total of 13 nights for this program during December 2009 --
June 2010, as specified in Table~\ref{tab:sample}.

\begin{table*}
\begin{minipage}{300mm}
{\tiny
\caption{Properties of the comparison stars.}
\label{tab:color}
\begin{tabular}{@{}ccc ccc ccc cc@{}} 
\hline \multicolumn{1}{c}{Object \footnote{* This source is not in
    SDSS so good information about its color is not available.}}
&\multicolumn{2}{c}{Star1(S1)} &\multicolumn{2}{c}{Star2(S2)}
&\multicolumn{3}{c}{R(mag)} &\multicolumn{3}{c}{$\delta(V-R)$}
\\ &\multicolumn{1}{c}{$\alpha_{2000.0}$} &
\multicolumn{1}{c}{$\delta_{2000.0}$}
&\multicolumn{1}{c}{$\alpha_{2000.0}$} &
\multicolumn{1}{c}{$\delta_{2000.0}$} &\multicolumn{1}{c}{Q}
&\multicolumn{1}{c}{S1} &\multicolumn{1}{c}{S2}
&\multicolumn{1}{c}{Q-S1} &\multicolumn{1}{c}{Q-S2}
&\multicolumn{1}{c}{S1-S2}\\ (1)&(2) &(3) &(4) &(5) &(6) &(7)&(8)&(9)
&(10) &(11) \\ \hline \hline WFM91 0226$-$1024$^*$
&02$^h$28$^m$31.17$^s$& $-$10$^d$17$^m$15.4$^s$&
02$^h$28$^m$40.65$^s$& $-$10$^d$15$^m$50.3$^s$&16.30&-----&-----&
----- &------ & ----- \\ J073739.96$+$384413.2 &07$^h$37$^m$22.23$^s$&
$+$38$^d$48$^m$55.5$^s$& 07$^h$37$^m$24.63$^s$&
$+$38$^d$41$^m$29.5$^s$&17.00&15.68&16.81& $-$0.01& $-$0.74& $-$0.73
\\ J084044.41$+$363327.8 &08$^h$40$^m$21.03$^s$&
$+$36$^d$37$^m$26.6$^s$& 08$^h$40$^m$32.32$^s$&
$+$36$^d$33$^m$34.5$^s$&16.59&16.32&16.17& $+$0.02& $-$0.12& $-$0.14
\\ \\ J084538.66$+$342043.6 &08$^h$45$^m$42.98$^s$&
$+$34$^d$17$^m$46.3$^s$& 08$^h$45$^m$45.14$^s$&
$+$34$^d$17$^m$07.3$^s$&16.95&16.05&16.66& $-$0.53& $-$1.20& $-$0.67
\\ J090924.01$+$000211.0 &09$^h$09$^m$17.27$^s$&
$+$00$^d$04$^m$01.3$^s$& 09$^h$09$^m$13.86$^s$&
$-$00$^d$01$^m$24.6$^s$&16.67&15.66&16.52& $-$0.82& $-$0.39& $+$0.43
\\ J094443.13$+$062507.4 &09$^h$44$^m$40.39$^s$&
$+$06$^d$28$^m$14.9$^s$& 09$^h$44$^m$34.10$^s$&
$+$06$^d$30$^m$33.0$^s$&16.24&15.62&16.76& $-$0.32& $-$1.19& $-$0.87
\\ \\ J094941.10$+$295519.2 &09$^h$49$^m$23.32$^s$&
$+$29$^d$54$^m$13.5$^s$& 09$^h$49$^m$42.06$^s$&
$+$30$^d$01$^m$07.1$^s$&16.06&15.80&15.44& $-$0.79& $-$0.93& $-$0.14
\\ J100711.81$+$053208.9 &10$^h$07$^m$17.82$^s$&
$+$05$^d$37$^m$05.1$^s$& 10$^h$07$^m$03.42$^s$&
$+$05$^d$34$^m$07.0$^s$&16.27&15.37&14.55& $-$0.18& $-$0.47& $-$0.29
\\ J111816.95$+$074558.1 &11$^h$18$^m$13.91$^s$&
$+$07$^d$46$^m$28.7$^s$& 11$^h$18$^m$05.79$^s$&
$+$07$^d$51$^m$21.5$^s$&16.15&15.94&15.59& $-$1.17& $-$0.49& $+$0.68
\\ \\ J112320.73$+$013747.4 &11$^h$23$^m$21.59$^s$&
$+$01$^d$45$^m$10.2$^s$& 11$^h$23$^m$30.68$^s$&
$+$01$^d$39$^m$55.6$^s$&15.84&15.98&15.67& $-$0.23& $-$0.57& $-$0.34
\\ J120051.52$+$350831.6 &12$^h$01$^m$13.18$^s$&
$+$35$^d$09$^m$03.9$^s$& 12$^h$01$^m$21.41$^s$&
$+$35$^d$04$^m$18.7$^s$&16.79&15.26&15.12& $-$0.40& $-$0.26& $+$0.14
\\ J120924.07$+$103612.0 &12$^h$09$^m$07.90$^s$&
$+$10$^d$34$^m$14.0$^s$& 12$^h$09$^m$16.06$^s$&
$+$10$^d$38$^m$38.0$^s$&16.50&15.82&15.07& $-$0.66& $-$1.08& $-$0.42
\\ \\ J123820.19$+$175039.1 &12$^h$38$^m$47.21$^s$&
$+$17$^d$56$^m$01.7$^s$& 12$^h$38$^m$19.22$^s$&
$+$17$^d$46$^m$07.9$^s$&16.42&15.62&15.15& $-$1.16& $-$0.27& $+$0.89
\\ J125659.92$+$042734.3 &12$^h$56$^m$47.73$^s$&
$+$04$^d$25$^m$25.1$^s$& 12$^h$56$^m$59.64$^s$&
$+$04$^d$31$^m$49.3$^s$&16.04&15.13&14.85& $-$0.34& $-$0.32& $-$0.02
\\ J151113.84$+$490557.4 &15$^h$11$^m$06.35$^s$&
$+$49$^d$08$^m$07.8$^s$& 15$^h$11$^m$34.13$^s$&
$+$49$^d$07$^m$17.0$^s$&16.49&16.07&16.12& $-$0.74& $-$0.19& $+$0.55
\\ \\ J152350.42$+$391405.2 &15$^h$23$^m$59.73$^s$&
$+$39$^d$17$^m$05.9$^s$& 15$^h$23$^m$51.32$^s$&
$+$39$^d$11$^m$48.4$^s$&16.65&16.35&15.64& $-$1.01& $-$0.11& $+$0.90
\\ J152553.89$+$513649.1 &15$^h$25$^m$57.63$^s$&
$+$51$^d$34$^m$51.9$^s$& 15$^h$26$^m$10.32$^s$&
$+$51$^d$36$^m$11.1$^s$&16.84&16.78&15.75& $-$1.03& $-$0.28& $+$0.75
\\ J154359.44$+$535903.2 &15$^h$44$^m$29.09$^s$&
$+$53$^d$58$^m$16.5$^s$& 15$^h$44$^m$19.22$^s$&
$+$53$^d$58$^m$03.2$^s$&17.05&16.48&16.06& $-$0.40& $-$0.48& $-$0.08
\\ J160207.68$+$380743.0 &16$^h$01$^m$34.59$^s$&
$+$38$^d$09$^m$31.4$^s$& 16$^h$01$^m$33.19$^s$&
$+$38$^d$05$^m$14.2$^s$&16.83&15.66&16.16& $-$0.47& $-$0.34& $+$0.13
\\
 
\hline
\end{tabular}                                                                                 
}
\end{minipage}
\end{table*} 

\subsection{Data Reduction} 
The raw photometric data was first pre-processed using standard
routines in the Image Reduction and Analysis Facility \footnote{IRAF
  is distributed by the National Optical Astronomy Observatories,
  which are operated by the Association of Universities for Research
  in Astronomy, Inc., under cooperative agreement with the National
  Science Foundation.} (IRAF) software. We generated a master bias
frame for the observing night by taking the median of all bias frames
taken on that night. This master bias frame was subtracted from all
the twilight sky flat image frames as well as from the source image
frames taken on that night. The routine step of dark frame subtraction
was not performed because the CCDs used in our observations were
cryogenically cooled to $-$120$^\circ$ C; at that temperature the
amount of thermal charge deposition is negligible for our brief
exposure times.  Then the master flat was generated by median
combining of several flat frames (usually more than 5 taken on the
twilight sky) in that passband. Next, the normalized master flat was
generated. Each source image frame was flat-fielded by dividing by the
normalized master flat in the respective band to remove pixel-to-pixel
inhomogeneities. Finally, cosmic ray removal was done from all source
image frames using the task \emph{cosmicrays} in IRAF.  \par

\subsection{Photometry}
The instrumental magnitudes of the comparison stars and the target
source are obtained from the data by using Dominion Astronomical
Observatory Photometry (DAOPHOT II) software to perform the concentric
circular aperture photometric technique (Stetson 1987, 1992). Aperture
photometry was carried out with four aperture radii, to wit, $\sim$
1$\times$FWHM, 2$\times$FWHM, 3$\times$FWHM and 4$\times$FWHM. Utmost
caution has been taken to deal with the seeing, and we have taken the
mean full width at half maximum $(FWHM)$ of 5 fairly bright stars on
each CCD frame in order to choose the apertures for the photometry of
that individual frame. The data reduced with different aperture radii
were found to be in good agreement. However, it was noticed that the
best S/N was almost always obtained with aperture radii of
2$\times$FWHM, so we adopted that aperture for our final analysis.
 
\section{Analysis}

\subsection{Selection of comparison stars}
\label{subs:stars}
The comparison stars are chosen on the basis of their proximity in
both location and magnitude to the quasar.  Preference was given to
those stars having magnitudes similar to that of the monitored quasar,
so that the errors in the Differential Light Curves (DLCs) will not be
dominated by any faint object (see Sect.~\ref{subs:stat}).  The
locations of the two best comparison stars for each BALQSO are given
in columns 2--5 of Table~\ref{tab:color}. \par

In addition, our atmosphere acts like a colour filter of variable
transparency, so the photometry of two stars (or quasar-star pair) of
different colours will be affected by different amounts because of
the changing air-mass during the monitoring (e.g.,  Eq.\ 2,
in Stalin et al.\ 2004). Therefore, for ascertaining the variability
properties from DLC, the colours of the two objects in the DLC really
should be similar. We list V$-$R colour differences for all pairs of
objects we observe in columns 9--11 of Table~\ref{tab:color}. For
most of the quasar-star pairs the V$-$R colour differences are
smaller than unity, except for seven pairs (out of a total of 38)
where the differences are the range of 1.0--1.20. Similarly, for the
star-star pairs, the colour differences for all pairs are smaller
than unity. Stalin et al.\ (2004) report a detailed investigation
quantifying the effect of colour differences, and they show that the
effect of colour differences of this amount on DLCs will be
negligible for a specific band (see also Carini et al.\ 1992).\par

We also used the star-star DLCs to identify any spikes in them
(unusually sharp rise or fall of the DLC over a single time bin),
assuming that the true star-star DLC is overall non-variable or at
worst reflects small stellar oscillations. Such spikes may arise from
improper removal of cosmic rays, cirrus clouds or some unknown
instrumental cause. Such outliers can sometimes significantly alter
the nominal statistics on short-term variations, especially when DLCs
do not have enough data points.  We typically have $\sim$30 individual
temporal data points in our sample; see Table~\ref{tab:res}, column 2.
We removed such outliers if they were more than 3$\sigma$ from the
mean, by applying a mean clip algorithm on the comparison star-star
DLCs. In cases where we find any such outliers we have censored those
time bin data points from our analysis of quasar-star DLCs as well.
Only DLCs once freed from any such outliers have been used for
carrying out our statistical analysis of microvariability. However, we
should stress that such outliers in our comparison star-star DLCs were
usually not present and never exceeded two data points.

\subsection{Statistics to  quantify microvariations}
\label{subs:stat}
\subsubsection{C-test statistics}
\label{subsubs:ctest}
To quantify microvariation of a DLC, by far the most commonly used
statistic is the so-called $C$-statistic (e.g., Jang \& Miller 1995;
Romero et al.\ 1999). This technique uses a variability parameter $C$,
which is an average of $C1$ and $C2$ with

\begin{equation}
  C1 = \frac{\sqrt{Var(q-s1)}}{\sqrt{Var(s1-s2)}} \hspace{0.2 cm} {\rm
  and} \hspace{0.2 cm} C2 = \frac{\sqrt{Var(q - s2)}}{\sqrt{Var(s1 - s2)}}.
\label{eq:cvalue}
\end{equation}

Here $Var (q-s1)$, $Var (q-s2)$ and $Var (s1-s2)$ are the variance of
observational scatters of the differential instrumental magnitudes of
the quasar$-$star1, quasar$-$s2 and star1$-$star2, respectively. The
normally adopted criterion to claim that variability is present is $C
\geq 2.576$, which corresponds to a nominal confidence level of $\ge
0.99$.\par

\subsubsection{F-test statistics}
\label{subsubs:ftest}
Despite the very common use of these $C$-statistics, de Diego
(2010) has pointed out it has severe problems.  Because it considers
the ratio of two standard deviations rather than of variances it does
not describe a normally distributed variable and it is not properly
centered with the mean expected value being zero; hence it is not a
good statistic and de Deigo (2010) concludes that the nominal critical
value for the presence of variability (i.e., 2.576) is usually too
conservative. \par

Another statistical method that can be used to quantify the presence
of microvariability is the $F$-test, which has been recently been
shown to be a more powerful and reliable tool for detecting
microvariability (de Diego 2010).   The $F$ value is computed as
\begin{equation}
 \label{eq.ftest}
 F_1=\frac{Var(q-s1)}{Var(s1-s2)} \nonumber \\
 F_2=\frac{Var(q-s2)}{ Var(s1-s2)},
\end{equation}
where $Var(q-s1)$, $Var(q-s2)$ and $Var(s1-s2)$ are the variances of
the quasar-star1, quasar-star2 and star1-star2 DLCs, respectively.
These $F$ values are then compared individually with the critical $F$
value, $F^{(\alpha)}_{\nu_{QS}\nu_{SS}}$, where $\alpha$ is the
significance level set for the test, and $\nu_{QS}$ and $\nu_{SS}$ are
the degrees of freedom of the quasar-star and star-star DLCs,
respectively.  The smaller the $\alpha$ value, the more improbable
that the result is produced by chance. Thus values of $\alpha=$
0.0001, 0.001 or 0.01 (the last assumed in our analysis) roughly
correspond to $5\sigma$, 3$\sigma$ or a 2.6$\sigma$ detections,
respectively. If $F$ is larger than the critical value, the null
hypothesis (i.e., no variability) is discarded.  Here we also note
that having two F-values, $F_1$ and $F_2$, allows us two choices in
deciding the presence of variability: (i) to take the average of $F_1$
and $F_2$, and compare it with critical $F$ value; (ii) to compare
$F_1$ and $F_2$ separately with the critical $F$ value. We prefer the
latter option, as it serves as further validation for the F-test; for
if one DLC indicates variability and one doesn't, these mixed signals
bring into question the reality of the putative variability. 

\subsubsection{Scaled F-test statistics}
\label{subsubs:fscaled}
Although the $F$-test is certainly better than the $C$-test, it should
be noted that for the $F$-test to give a truly reliable result the
error due to random noise in the quasar--star and star-star DLCs
should be of a similar order, apart from any additional scatter in the
quasar-star DLC due to possible QSO variability. For instance, if both
comparison stars are either brighter (fainter) than the monitored
quasar, then a false alarm detection (non-detection) is possible due
to the very small (large) photon noise variance of the star-star DLC
compared to the quasar-star DLCs. This in practice can happen, as
sometimes it is difficult to fulfill the desiderata of having
non-variable comparison stars within the quasar CCD image frame that
are very similar in magnitude to the QSO.\par

In our sample we have tried to choose non-variable comparison stars
in proximity to the magnitude of the quasar (see
Sect.~\ref{subs:stars}), but it was not possible to fulfill this
requirement for all quasars. So sometimes in performing the $F$-test
we may have to compare the variance of star-star DLCs involving stars
substantially brighter than the quasar, where scatter due to photon
noise is very small, with the noisier quasar-star DLC. In such cases,
the standard statistics of the $F$-test do seem to give too much
weight to even very nominal fluctuations in a quasar-star DLC.  A
sensible way to deal with this real problem is to scale the star-star
variance by a factor, $\kappa$, which is proportional to the ratio of
the noise in the quasar-star and star-star DLCs. One logical choice
along these lines is to consider the ratio of the average squared
error in the quasar-star and star-star DLCs i.e., 
\begin{equation}
\kappa=\left[\displaystyle{\frac{\sum_\mathbf{i=0}^{N}\sigma^2_{i,err}(q-s)/N}{\sum_{i=0}^{N}\sigma^2_{i,err}(s1-s2)/N}}\right] \equiv \frac{\langle\sigma^2(q-s)\rangle}{\langle\sigma^2(s1-s2)\rangle},
\label{eq:kappa}
\end{equation}
where $\sigma^2_{i,err}(q-s)$ and $\sigma^2_{i,err}(s1-s2)$ are,
respectively, the errors on individual points of the quasar-star and
star-star DLCs, as returned by the DAOPHOT/IRAF routine. Then the
scaled F-value, $F^{s}$, can be computed as,
\begin{equation}
\label{eq.fstest}
 F_{1}^{s}=\frac{Var(q-s1)}{\kappa Var(s1-s2)}, \nonumber \\
 F_{2}^{s}=\frac{Var(q-s2)}{\kappa Var(s1-s2)}.
\end{equation}

Here scaling the variance of the star-star DLC by $\kappa$ basically
amounts to normalizing the variance of the DLC by the mean of the
squared errors of its individual points (i.e., by
$\langle\sigma^2\rangle$ in Eq.~\ref{eq:kappa}).  This is sensible, as
we know that for no intrinsic variability present in a light curve,
the variance gives an estimate of the square of the mean errors (i.e.,
$\langle\sigma^2\rangle$) of the light curve. These
$\langle\sigma^2\rangle$s of the light curves depend on the
brightnesses of the observed objects, so to remove any effects of
brightness on the variances (used in the standard F-test as
Var(q$-$s1)/Var(s1$-$s2)) of light curves it should be better to use
the variances that have been normalised by their
$\langle\sigma^2\rangle$ values.

The value of scale factor, $\kappa$, used in the {\it scaled} F-test
(Eq.~\ref{eq.fstest}) will be near unity if the quasar and stars are
of similar magnitude, and as a result it will give a similar F-value
as is given by the standard F-test (see Eq.~\ref{eq.ftest}).  On the
other hand, if both comparison stars are either brighter (fainter)
than the monitored quasar, then $\kappa$ will be larger (smaller) than
unity.  As a result, $\kappa$ will reasonably scale the variance of
comparison star-star DLCs for any magnitude difference between stars
and quasar, and hence avoid the problem with the standard $F$-test
which does seem to give too much weight to even very nominal
fluctuations in a quasar$-$star DLC, when it is compared to brighter
star$-$star DLCs. \par

Other alternatives to the standard F-test are the use of one-way
analysis of variance (ANOVA) or a $\chi^2$ test (e.g., de Diego
2010).  For an appropriate use of ANOVA the number of data points in
the DLC needs to be large enough so as to have many points in each
subgroup used for the analysis; however, this is not possible for our
observations as we typically have only around 30 data points in our
light curves.  For the appropriate use of a $\chi^2$ test, the errors
of individual data points need to have Gaussian distributions and
those errors should be accurately estimated.  It has been claimed in
the literature that errors returned by photometric reduction routines
in IRAF and DAOPHOT usually are underestimated, often by factors of
1.3--1.75 (Gopal-Krishna et al.\ 2003; Sagar et al.\ 2004; Bachev et
al.\ 2005), which makes the use of a $\chi^2$ test less desirable for
such real photometric light curves.  However, as our scale factor
depends on the ratio of average squared errors, this possible caveat
does not affect our scaled $F$-test analysis.\par

In conclusion, we propose that by applying such scaling to the
variance of the star-star DLC we can perform a {\it scaled} $F$-test,
where our scale factor is designed so that it: (i) takes care of the
difference in magnitude between the QSO and star in quasar-star and
star-star DLCs; (ii) retains the requirement that both the variance
being compared in a $F$-test should have a $\chi^2$ distribution,
which is not the case in $C$-statistics; and (iii) cancels out the
problem of uncertain error underestimation by DAOPHOT/IRAF routines
reported by many other authors, in that our scaling factors depend on
ratios of averaged squared errors.  Therefore we report our final
results based on this scaled $F$-test. However it is also worthwhile
to compute $C$-values and standard $F$-values to facilitate the
comparison of results for variability based on them with those based
on our newly proposed {\it scaled} $F$-test.

\begin{table*}
 \centering
 \begin{minipage}{500mm}
{\tiny
\caption{Microvariability observations of BALQSOs.}
\label{tab:res}
\begin{tabular}{@{}ccc ccc ccc cc cc@{}} 
\hline \hline \multicolumn{1}{c}{QSO} &{N} &{T} &{C-test}
&\multicolumn{4}{c}{F-test}
&\multicolumn{3}{c}{Variability?{\footnote{V=variable, i.e., confidence
      $\ge 0.99$; Pv=probable variable, i.e., $0.95-0.99$ confidence; Nv
      =non-variable, i.e., confidence $< 0.95$.\\
Variability status values based on quasar-star1 and quasar-star2 pairs are separated by a comma.}}}
&{$\sqrt\kappa${\footnote{Here
       $\kappa=\langle\sigma^2(q-s)\rangle/\langle\sigma^2(s1-s2)\rangle$ (as in Eq.~\ref{eq:kappa}), is
 used to scale the variance of star-star DLCs for  the scaled F-test.}}}
&{$\sqrt { \langle \sigma^2_{i,err} \rangle}$}\\ & &{(hr)}&C-value
&{$F_{1}$},{$F_{2}$}&{$F_{1}^{s}$},{$F_{2}^{s}$}&{$F_{c}(0.95)$}&{$F_{c}(0.99)$} 
&{C-test}&F-test &F$_s$-test &$\frac{}{}$ &(Q-S)\\ (1)&(2) &(3) &(4) &(5) &(6)
&(7)&(8)&(9) &(10) &(11) &(12)&(13) \\ \hline \hline 
WFM91 0226$-$1024     & 31 & 4.04 &2.20, 2.47 & 4.84, 6.10& 1.80, 2.25& 1.84 & 2.39  & Pv,Pv&  V,V&Nv,Pv& 1.64 & 0.02\\ 
J073739.96$+$384413.2 & 40 & 5.54 &1.45, 1.83 & 2.11, 3.35& 1.73, 1.78& 1.70 & 2.14  & Nv,Nv& Pv,V&Pv,Pv& 1.24 & 0.01 \\ 
J084044.41$+$363327.8 & 33 & 3.86 &1.24, 0.84 & 1.54, 0.71& 1.02, 0.51& 1.80 & 2.32  & Nv,Nv&Nv,Nv&Nv,Nv& 1.21 & 0.01\\ \\ 
J084538.66$+$342043.6 & 32 & 3.90 &1.89, 1.99 & 3.57, 3.96& 1.07, 1.08& 1.82 & 2.35  & Nv,Pv&  V,V&Nv,Nv& 1.87 & 0.01 \\ 
J090924.01$+$000211.0 & 33 & 3.50 &1.74, 1.56 & 3.04, 2.44& 2.35, 1.25& 1.80 & 2.32  & Nv,Nv&  V,V& V,Nv& 1.27 & 0.02\\ 
J094443.13$+$062507.4 & 25 & 2.79 &1.37, 1.65 & 1.87, 2.71& 2.80, 2.05& 1.98 & 2.66  & Nv,Nv& Nv,V& V,Pv& 0.98 & 0.01 \\ \\ 
J094941.10$+$295519.2 & 36 & 3.50 &2.17, 1.82 & 4.71, 3.31& 2.56, 1.94& 1.76 & 2.23  & Pv,Nv&  V,V& V,Pv& 1.33 & 0.01\\ 
J100711.81$+$053208.9 & 33 & 3.88 &1.49, 1.80 & 2.23, 3.25& 0.46, 0.75& 1.80 & 2.32  & Nv,Nv& Pv,V&Nv,Nv& 2.14 & 0.04 \\ 
J111816.95$+$074558.1 & 36 & 3.74 &1.56, 1.73 & 2.44, 3.00& 1.77, 2.58& 1.76 & 2.23  & Nv,Nv&  V,V& Pv,V& 1.13 & 0.01\\ \\ 
J112320.73$+$013747.4 & 42 & 3.48 &1.17, 1.21 & 1.38, 1.47& 1.12, 1.47& 1.68 & 2.09  & Nv,Nv&Nv,Nv&Nv,Nv& 1.05 & 0.01 \\ 
J120051.52$+$350831.6 & 30 & 3.86 &3.25, 3.24 &10.57,10.50& 1.64, 1.64& 1.86 & 2.42  &   V,V&  V,V&Nv,Nv& 2.53 & 0.02\\ 
J120924.07$+$103612.0 & 39 & 4.10 &3.37, 3.06 &11.36, 9.34& 2.85, 2.41& 1.72 & 2.16  &   V,V&  V,V&  V,V& 1.98 & 0.02 \\ \\ 
J123820.19$+$175039.1 & 30 & 3.77 &1.97, 1.98 & 3.87, 3.92& 0.81, 0.74& 1.86 & 2.42  & Pv,Pv&  V,V&Nv,Nv& 2.24 & 0.01\\ 
J125659.92$+$042734.3 & 33 & 3.84 &2.32, 2.20 & 5.40, 4.84& 2.99, 2.68& 1.80 & 2.32  & Pv,Pv&  V,V&  V,V& 1.34 & 0.01 \\ 
J151113.84$+$490557.4 & 66 & 3.70 &1.36, 1.37 & 1.84, 1.88& 1.37, 1.79& 1.51 & 1.79  & Nv,Nv&  V,V& Nv,V& 1.09 & 0.01\\ \\ 
J152350.42$+$391405.2 & 43 & 2.75 &1.94, 1.83 & 3.75, 3.35& 1.97, 1.96& 1.67 & 2.08  & Nv,Nv&  V,V&Pv,Pv& 1.34 & 0.01 \\ 
J152553.89$+$513649.1 & 24 & 3.00 &1.16, 1.12 & 1.35, 1.25& 0.73, 1.05& 2.01 & 2.72  & Nv,Nv&Nv,Nv&Nv,Nv& 1.22 & 0.01\\ 
J154359.44$+$535903.2 & 25 & 2.90 &1.94, 2.01 & 3.75, 4.05& 1.33, 1.64& 1.98 & 2.66  & Nv,Pv&  V,V&Nv,Nv& 1.62 & 0.02 \\ 
J160207.68$+$380743.0 & 32 & 4.07 &2.12, 2.34 & 4.50, 5.48& 2.01, 2.08& 1.82 & 2.35  & Pv,Pv&  V,V&Pv,Pv& 1.56 & 0.01 \\ 
\hline
\end{tabular}  
}  
\end{minipage}
\end{table*}

\section{Results}
\label{sec:res}
\subsection{Differential light curves (DLCs)} 
\label{subsec:dlc}
The R-band differential light curves (DLCs) of our sample are shown in
Figures~\ref{fig:s01to04}-\ref{fig:s11to19}.  For each quasar the
upper panel gives the star-star DLCs of the two best comparison stars
and the two lower panels give the quasar-star DLCs.  We first give
brief notes on each individual source and then present our results
based on all three statistical tests, the $C$-test, the {\it standard}
$F$-test and the {\it scaled $F$-test}. As discussed in the previous
section our final results will be based on scaled $F$-test but the
results based on the other two tests will facilitate their
intercomparisons, allowing us to discuss their relative merits.

\subsection{Brief notes on individual sources}
\subsubsection{[WFM91] 0226$-$1024}
[WFM91] 0226$-$1024 is a high ionization BAL (HiBAL) QSO, having
balnicity index (BI) =7344 km s$^{-1}$, and detachment index (DI) =
4.72 km s$^{-1}$ (Weymann et al.\ 1991).  This BALQSO was reported as
a normal QSO and earlier photometric monitoring to search for its
microvariability over $\sim$3.4 hr did not yield any positive
microvariability detection (Bachev et al.\ 2005).  Over our
observational run of $\sim$ 4 hr a peak in the DLC is apparent by
visual inspection.  However, due to the rather high error bars in the
DLCs, the $C$-test and the $F$-test have indicated this source is a
possible variable and variable, respectively. But our scaled $F$-test
shows it is not variable; nonetheless as the scaled $F$-test with one
standard star has shown it as possible variable, this BALQSO is a
prime candidate for additional monitoring.  For the remainder of the
sources we will not discuss the details of the differences between the
different statistical tests, reserving a general discussion for the
next subsection.

\subsubsection{J073739.96$+$384413.2}
J073739.96$+$384413.2 is a Low ionization (LoBAL) QSO. We have
monitored this source over a span of more than $\sim$ 5
hr. Statistical analysis of its DLC shows it is a probably variable
source.

\subsubsection{J084044.41$+$363327.8}
Becker et al.\ (1997) reported the discovery of this unusual LoBAL
QSO.  A spectropolarimetry study by Brotherton et al.\ (1997) reveals
that it is a highly polarized BALQSO, with the continuum polarization
rising steeply toward shorter wavelengths, while keeping a constant
position angle in the continuum.  This source was observed for
$\sim$3.8 hours, but no evidence of microvariations were detected in
its DLC.

\subsubsection{J084538.66$+$342043.6}
The source, also known as CSO230, has a black hole with mass
$16.4\times10^9M_{\odot}$ , estimated using its H$\beta$ broad line
width (e.g., Yuan et al.\ 2003). It is HiBAL QSO having balnicity
index of $2564\pm 1.17$ km s$^{-1}$, and absorption index (AI) of
$4091\pm1.28$ km s$^{-1}$ (Trump et al.\ 2006). This source has been
extensively studied spectroscopically.  Barlow et al.\ (1992) has
studied its spectral variability during four epochs over a 17-month
time span. They found three distinct levels in the broad absorption
lines of Si IV 1397\AA~ and C IV 1549\AA~.  A broad-band monitoring
effort during this period showed that the continuum level remained
constant to within 10 percent. The source remained non-variable during
our observational run of $\sim$4 hr.

\subsubsection{J090924.01$+$000211.0} 
This is a HiBAL, with balnicity index of 71$\pm0.90$ km s$^{-1}$(Trump
et al.\ 2006).  This is binary quasar system (Hennawi et al.\ 2006).
We observed this source for $\sim3.5$ hr. Statistical analysis of its
DLC does not give a good indication of rapid variability according to
the scaled F-test.

\subsubsection{J094443.13$+$062507.4}
This is LoBAL quasar with balnicity index of 820$\pm0.53$ km
s$^{-1}$(Trump et al. \ 2006). This source was found to be probably
variable during the course of our $\sim$ 2.7 hr observation, which
makes it a potential source for further microvariability study.

\subsubsection{J094941.10$+$295519.2}
This source is a prime candidate for microvariability and an intensive
search over a long time span (from 1993--1996) was performed by
Gopal-Krishna et al.\ (2000) in their programme to search for
intranight optical variability in RQQSOs.  They found evidence of an
$\sim$0.05 mag probable variation and marginal evidence of $\sim$0.03
mag variation over observations lasting 2.5 hr and 4.5 hr,
respectively. In addition, Jang et al.\ (2005) monitored this source
for two nights for durations of 3.9 hr and 2.0 hr respectively, but
they did not find any sign of variability.  Our scaled $F$-test
analysis indicates that it possibly exhibited microvariability during
our observation lasting $\sim$ 3.5 hr.

\subsubsection{J100711.81$+$053208.9}
This is a HiBAL QSO with balnicity index of 2901$\pm0.88$ km s$^{-1}$
(Trump et al.\ 2006).  For this source the quasar-star DLC is a
noisier than usual. We did not find any signature of microvariability
in its DLC during our observation of $\sim$ 3.8 hr.  To reach a firmer
conclusion as to its rapid variability this source merits additional
observations.

\subsubsection{J111816.95$+$074558.1}
PG 11514$+$081, also known as the ``triple quasar'', was the second
gravitational lens found (Kristian et al.\ 1993), to have three
components with identical spectra (at a redshift of 1.722). Hubble
Space Telescope observations resolved the system PG 11514$+$081 into
four point sources and a red extended lens galaxy (Kristian et
al.\ 1993). The source was observed for $\sim$ 3.7 hr and found to be
a probably variable source, which makes it an excellent candidate for
future microvariability investigations.

\subsubsection{J112320.73$+$013747.4}
Meylan and Djorgovski (1989) reported that this quasar is probably
lensed by a galaxy at z$\sim0.6$.  The UV line profile structure found
with the International Ultraviolet Explorer in this gravitational lens
candidate indicates pronounced BAL structure in the high-ionization
resonance lines of O VI 1033\AA~ and N V 1240\AA.  Michalitsianos et
al.\ (1997) performed a comparison of far-UV spectra, with data
separated by nearly 10 months, that indicated that changes occurred in
both absorption and ionization levels associated with BAL structure in
the QSO.  We found this source to be non-variable during our $\sim$
3.4 hr observation.

\subsubsection{J120051.52$+$350831.6}
This source is a HiBAL with a balnicity index of 4600$\pm2.48$ km
s$^{-1}$ (Lamy et al.\ 2004).  This source did not show any
significant microvariation over an observational run of $\sim$3.8 hr
and is non-variable according to the scaled $F$-test.  Although the
$C$-statistic showed it as a strong contender to have presented
microvariability, that result appears to have been induced because of
its relatively bright comparison stars, as discussed above.

\subsubsection{J120924.07$+$103612.0} 
Significant variations were noticed in the DLC over our observational
run of $\sim$ 4 hr.  Note that a coherent variability trend can be
seen in both the quasar--star DLCs.  Statistical analyses using the
$C$-test, $F-$test and scaled $F-$test all strongly indicate the
presence of microvariability.

\subsubsection{J123820.19$+$175039.1}
This LoBAL is in the Large Bright Quasar Survey, and was also detected
in the Chandra BAL quasar survey (Green et al.\ 2001).  We did not
find any signature of microvariation in its DLC over an observational
period of $\sim$ 3.77 hr.

\subsubsection{J125659.92$+$042734.3}
This source has been extensively studied for optical microvariability.
Barbieri et al.\ (1984) did not find any signature of variability in
their observations. In their search for intranight optical variability
in RQQSOs.  Gopal-Krishna et al.\ (2000) observed this source twice
for 5 hr each time and on one of those nights, during which they had
unfortunately sparse sampling, saw a hint of microvariation.  We have
investigated this source for $\sim$ 3.8 hr, and the statistical
analysis of its DLCs showed clear evidence of microvariability.

\subsubsection{J151113.84$+$490557.4}
This LoBAL quasar has a balnicity index of 802$\pm1.33$ km s$^{-1}$
(Trump et al.\ 2006).  We observed this source for $\sim$ 3.5 hr but
found no  overall evidence of microvariability, although one star-QSO DLC
was nominally variable.

\subsubsection{J152350.42$+$391405.2}
This is a LoBAL QSO having a balnicity index of 7147$\pm1.66$ km
s$^{-1}$ (Trump et al.\ 2006).  This bright quasar was found in the
third Hamburg Quasar Survey (Hagen et al.\ 1999).  This source
appeared to be variable in a 20 cm radio study (Becker et al.\ 2000).
We found it to be probably variable over the course of an observing
run of $2.7$ hr.

\subsubsection{J152553.89$+$513649.1} 
This source, also known as CSO 755, is a strongly polarized ($\sim3.9$
per cent) BALQSO (Glenn et al.\ 1994).  A strongly polarized continuum
and unpolarized emission lines indicate that its polarization arises
by scattering very near the central source (Glenn et
al.\ 1994). XMM-Newton spectroscopy of this luminous quasar gives a
photon index of $\Gamma =1.83^{+0.07}_{-0.06}$ and a flat (X-ray
bright) intrinsic optical-X-ray spectral slope of $\alpha _{ox}=-1.51$
(Shemmer et al.\ 2005).  The source shows evidence for intrinsic
absorption, having a column density of N(H) $\sim 1.2 \times 10^{22}$
cm$^{-2}$.  This is among the lowest X-ray columns measured for a
BALQSO (Shemmer et al. 2005).  We detected no signature of
microvariability over a short run of $\sim$ 2.9 hr.

\subsubsection{J154359.44$+$535903.2}
\label{notes:J154359}
J154359.3$+$535903 is also known as SBS 1542$+$541 as this source was
discovered in the Second Byurakan Survey (Stepanyan et al.\ 1991).  It
has many interesting properties: its BAL has a very high degree of
ionization (Telfer et al.\ 1998), an associated absorption system and
damped Ly$\alpha$ (DLA) absorption system, and a strong X-ray
absorption (Green et al.\ 2001).  This bright high-redshift HiBAL QSO
`has very highly ionized BALs (including O VI, Ne VIII, and Si XII;
Telfer et al.\ 1998) and appears to have an X-ray brightness typical
for a non-BAL of its optical luminosity. Bechtold et al.\ (2002) has
found intervening metal absorption systems at $z$ = 1.41, 0.1558, and
0.72 along its line of sight. We found this source to be non-variable
during our observation of $\sim 4$ hr.

\subsubsection{J160207.68$+$380743.0}
This source was continuously observed for $\sim3.7$ hr. We found 
this as a  probably variable source, which makes this source   another 
potentially good candidate for microvariability  studies in the future. 

\subsection{Variability results based on different statistical test}
\label{subsec:resvar}
The results of our analysis are summarized in Table~\ref{tab:res}; we
applied both the $C$-statistic and the scaled $F$-test, as discussed
above (e.g., see Sect~\ref{subs:stat}). In the first three columns we
list the object name, number of data points ($N_{points}$) used in the
DLC and the duration of our observation.  The fourth column lists the
pair of $C$-values based on star1 and star2 (Eq.~\ref{eq:cvalue})
while the fifth and sixth columns list the pair of $F$-values in the
standard and scaled $F$-test. Columns 7 and 8, respectively, give
$F_{c}$ for 0.95 and 0.99 confidence levels. Columns 9, 10 and 11
respectively, list the pairs of variability statuses using star1 and
star2 based on $C$-statistics, the standard $F$-test and the scaled
$F$-test. The status, based on both star1 and star2 are listed
separately rather than using their average value so as to impose as
additional validation: is the variability status based on individual
stars are consistent with one another or not?  In these pairs of
variable status indicators using a quasar-star DLC, the quasar is
marked as variable (`V') for a $C$-value $\ge 2.576$ or $F$-value $\ge
F_{c}(0.99)$, which corresponds to a confidence level $\ge 0.99$.  The
quasar is marked as `probably variable' (Pv) if the $C$ value of
quasar-star DLC is in the range 1.950 to 2.576 or if the $F$-value is
between $F_{c}(0.95)$ and $F_{c}(0.99)$.  Those sources for which the
$C$-values are less than 1.95, or the $F$-value are less than
$F_{c}(0.95)$ are marked Non-variable (`Nv').  Column 12 lists the
square root of scaling factor, $\sqrt\kappa$, where it is computed by
$\kappa=\langle\sigma^2(q-s)\rangle/\langle\sigma^2(s1-s2)\rangle$ (as
in Eq.~\ref{eq:kappa}), and has been used to scale the variance of the
star-star DLCs while computing the $F$-value in the
scaled-$F$-test. The last column gives our photometric accuracy,
{$\sqrt { \langle \sigma^2_{i,err} \rangle}$ } in the quasar-star
DLCs, which typically are between 0.01$-$0.02mag.\par

 As can be seen from Columns 9 -- 11 of Table~\ref{tab:res} the
 variability status indicators based on quasar-star1 and quasar-star2
 are often not consistent with one another.  The importance of our
 choice to mark the variable status separately based on individual
 star vs quasar DLCs can be illustrated by taking the example of
 J094941.10+295519.2.  Based on the $C$-test its DLC with respect to
 star1 shows it as a probable variable but with star2 as
 non-variable. The standard $F$-test terms it as variable based on
 both star1 and star2 DLCs.  However, this QSO's status using the
 scaled $F$-test is variable based on the first star and probably
 variable using the second star.  The average of the scaled $F$-value
 for this source comes out to be 2.25, which is just above the
 critical $F$-value of 2.23 for 0.99 confidence, and hence it would be
 classified as a variable source if we used that average
 criterion. However, an examination the DLC of this source in top
 right panel of Fig.~\ref{fig:s05to10} by eye indicates that there is
 no variation that can defined coherently by more than 2
 points. Therefore, to exclude such questionable variability and to be
 on the conservative side for unambiguous microvariable detection,
 only those sources should be termed as variable for which both
 quasar-star1 and quasar-star2 DLCs mark the source as variable (i.e
 `V,V' in Table~\ref{tab:res}).  Probably variable sources are taken
 as those for which either both the status are of probable variable
 (i.e., `Pv,Pv' in Table~\ref{tab:res}) or one quasar-star DLC marks
 it as a probable variable and the other as a variable (i.e., `Pv,V'
 or `Pv,V' in Table~\ref{tab:res}). Sources termed as non-variable
 (`Nv') are those for which at least one of the status based on
 quasar-star1 and quasar-star2 DLC marked them as non-variable (i.e.,
 at least one `Nv' status in Table~\ref{tab:res}). \par

  Column (9) of Table~\ref{tab:res} indicates that the $C$-statistics
  shows two sources as variable and four as probably variable. The
  scaled $F$-test shows two sources as variable and six sources as
  probably variable. As we have discussed above (in
  section~\ref{subsubs:fscaled}), that scaled $F$-test is better for
  our work (and probably also better in many observations made by
  others) than the standard $F$-test due to differences between the
  magnitudes of the quasars and their comparison stars. This is also
  evident from column (10) of Table~\ref{tab:res} which shows that the
  standard $F$-test would give 13 sources as variable and two as a
  probably variable, indicating that this test certainly suffers from
  the problem related to small variances of the brighter star-star
  DLCs, at least for our sample of BALQSOs.

 Although the $C$-test and the {\it scaled} $F$-test both give two
 sources as variable, it is not difficult to appreciate the scaled
 $F$-test merits over the $C$-test by taking specific examples. For
 instance, J120051.52$+$350831.6 is a BALQSOs with a $C$-value of 3.25
 from quasar-star1 DLC and 3.24 from quasar-star2 DLC, which seems to
 make it a clear case of INOV detection, particularly since the
 $C$-statistic is usually conservative.  However, by looking at the
 DLCs for this source in the top left panel of Fig.~\ref{fig:s11to19},
 it is clear even by eye that: (i) there is likely to have been a
 random fluctuation (not a coherent one) for the last 9 points of the
 DLCs; and (ii) its comparison stars are about 1.5 mag brighter than
 this quasar, which makes the variance of the star-star DLC very small
 (due to small photon noise). As a result the $C$-value will be
 artificially very high, leading to false detections. This flaw also
 crops up in the standard $F$-test, but is eliminated in the scaled
 $F$-test which termed this BALQSO as a non-variable source (not even
 probably variable). Another source, J120924.07$+$103612.0, has a
 $C$-value of 3.37 from the quasar-star1 DLC and 3.06 from the
 quasar-star2 DLC but these are probably so high because of the
 $\sim$1.2 mag brighter comparison stars; however, this BALQSO also
 shows a coherent variability trend (even by eye), and is also termed
 as variable by the scaled $F$-test.  These empirical examples, and
 the fact that the scaled $F$-test detects two cases of unambiguous
 variability in comparison to the $C$-test which makes only one
 unambiguous detections (after eliminating the false positive case
 mentioned above) clearly shows that the scaled $F$-test, beside being
 more sensitive than the $C$-test to small amplitude variability, is
 also sufficiently robust to eliminate nearly any false alarm
 detections.
   
Therefore, finally, we rely on the result given by the scaled
$F$-test, by which we find two unambiguous detections of
microvariability in our sample of 19 BALQSOs up to an accuracy of
0.01-0.02mag (see columns 9,10, 11 and 13 of Table~\ref{tab:res}). As
a result, our sample shows that about 10-11 per cent of BALQSOs (i.e.,
2 out of 19 sources) certainly showed microvariability (at a
confidence level of 0.99).

\section{Discussion and Conclusions}
As noted in the introduction, there have been rather extensive
examinations of the frequency of optical microvariability for RQQSOs
as well as blazars and other RLQSOs. The typical duty cycle (DC) for
blazars is 60--65 per cent (e.g., Gupta et al.\ 2005), while for
normal quasars it has been found to be around 20--25 per cent (e.g.,
Carini et al.\ 2007). For both these classes the number of sources in
each total sample was quite large, so these values should be
reasonably reliable, and support the hypothesis that most of these
rapid variations arise, or at least are amplified, in the relativistic
jets (e.g., Jang \& Miller 1995; Gopal-Krishna et al.\ 2003).  The
interesting class of radio-quiet BALQSOs was reported to have a 50 per
cent DC but this sample had only 6 members (Carini et al.\ 2007).
Therefore, one of the reasons for the difference in DC results could
be poor statistics in the previous study and better statistics now
with a sample about a factor of three larger.  Apart from sample size,
some of the difference might be due to differences in the typical
length of the observation.  As long known, lengthier observations of
blazars are more likely to reveal variability (e.g., Carini 1990),
which was also later shown to be the case for RQQSOs (Gupta et
al.\ 2005, Carini et al.\ 2007).  Carini et al.\ (2007) found that
RQQSOs that were monitored for about 6-7 hr showed the highest
fraction of microvariability, typically around 24 per cent. In the 4
hour observation range, which is where most of our observations fall,
less than 10 percent of sources were found to have microvariability.
Therefore the fact that we found about 10-11\% DC for radio-quiet
BALQSOs is in agreement with the results from the literature for other
radio-quiet non-BALQSOs indicates that the BAL nature does not have an
effect on the presence of microvariability.

In addition, as we discussed in Sections~\ref{subs:stat} and
\ref{subsec:resvar}, this DC fraction also depends on what statistical
test has been used to decide on the significance of microvariation
(see de Diego 2010 for details). Most of these previous studies have
used the $C$-test, which has been shown recently to be an unreliable
and usually too conservative method to detect microvariation, when
compared to proper statistics such as the $F$-test (de Diego 2010).
So, to allow comparison with earlier results, we also computed duty
cycles (DCs) of BALQSOs in our sample using the $C$-test, which shows
only one out of 19 source as variable (excluding one false detection,
as discussed in Sec.~\ref{subsec:resvar}), resulting in a DC of about
5 per cent.  As a result the DCs of all classes of AGN may increase if
their DLCs are analyzed with the scaled $F$-test rather than with the
usually more conservative $C$-test, which, for the 4 hour observation
range, were reported at less than 10 percent for RQQSOs (Carini et
al.\ 2007). Therefore, after taking into account the observation
length and the dependence on statistical test used, it seems that the
DC of radio-quiet BALQSOs is likely to be of a similar value to the DC
of non-BAL RQQSOs, but without redoing all past analyzes with the
scaled $F$-test we cannot be certain of this assertion. Apart from
this comparison using the $C$-test, all our final results and
conclusions are based on the more reliable scaled $F$-test (see
Sect.~\ref{subsec:resvar}), which gives our new result of an
approximately 11 per cent DC for radio-quiet BALQSOs.\par 

The phenomenon of microvariability was first noticed for blazars, and
for them microvariability almost certainly arises from a relativistic
jet.  However, given the lower DCs for RQ AGN it is still unclear if
the nature of intranight variability is the same in these objects, or
if it arises from processes in the accretion disc itself (e.g.,
Mangalam \& Wiita 1993; Chakrabarti \& Wiita 1993), and thus could
possibly be used to probe the accretion disc (e.g., Gopal-Krishna et
al.\ 2000 and references therein).  Recent modeling suggests that even
for RQQSOs, jet based models should be the most efficient way to
produce microvariability (e.g., Czerny et al.\ 2008).  Such jet-based
models should predict a difference in equivalent widths of emission
lines of variable and non-variable sources, as they should be smaller
in the former, due to its dilution by jet components.  However,
recently this hypothesis was shown to be unlikely based on an analysis
of spectra of a set of RQQSOs that had already been searched for
microvariability (Chand et al.\ 2010).  Some other possibilities
include disc based models where variation can be due to: variable hard
radiation from near the disc center that is reprocessed into the
UV-optical region (e.g., Ulrich et al. 1997); instabilities in the
accretion flow itself producing multiple hot-spots (e.g., Mangalam \&
Wiita 1993); disco-seismological modes within the disc (e.g. Nowak \&
Wagoner 1991, 1992).  In the first case above the variable hard
radiation instead may also come from a hot corona above the accretion
disc (Merloni \& Fabian 2001).\par

The shortest variability time scale in this scenario can be associated
with the light-crossing time, which will be larger for higher central
black hole masses (e.g see Bachev et al.\ 2005).  In a scenario with
instabilities in an accretion flow, if one assumes that the inner part
of the flow operates through an optically thin advective mode, then
the border between it and outer thin disc may be good candidate for
the region where these instabilities may occur (e.g., Gracia et
al.\ 2003; Krishan, Ramadurai \& Wiita 2003).  The time scale may also
be associated with the much longer accretion time scale.
Observational estimation of such variability time scales could
possibly be used to distinguish between above various disc based
scenarios. However, for luminous QSOs such as those in our sample that
should have black hole masses in excess of $10^8$ M$_{\odot}$, even
the fastest disc based time scale can be a few hours. As our
monitoring of each source rarely exceeded 4 hours and was sometimes
unevenly sampled, it is not possible for us to obtain reliable
estimates of the variability time scales for the minority of variable
sources.  Therefore we are not able to distinguish between the various
disc based scenarios mentioned above, nor can we cleanly distinguish
between jet based and disc based models.\par

Our larger sample (a factor of three improvement) of radio-quiet
BALQSOs, aided by more robust detection criteria, have allowed us to
conclude that the fraction of radio-quiet BALQSO showing
microvariability are about 11\% for an observation length of about 4
hr. This new DC of 11\% is similar to the usual low microvariability
fraction of normal RQQSOs with observation length similar to our
observation, though we note those DCs were obtained using the
$C$-statistic and not our scaled $F$-test. This similarity in
microvariability frequency provides some support for models where
radio-quiet BALQSO do not appear to be a special case of the RQQSOs.\par

Further extension this type of study to radio-loud BALQSOs will be
important in obtaining good values for the microvariability percentage
for all types of BALQSOs.  Such additional observations will also help
in understanding that whether we are viewing BALQSOs closer to the
disc plane or to the perpendicular to the disc.  In the latter case
higher DCs are expected assuming that the cause of microvariation is
related to the relativistic jet. In addition, to investigate whether
and how X-ray and optical microvariability are correlated in BALQSOs,
it will be useful to carry out future simultaneous multi-color optical
and X-ray monitoring observations (e.g., Ram{\'{\i}}rez et al. 2010).
Finally, to ascertain whether or not such microvariation depends on
spectral properties, additional optical and X-ray spectral analyses of
the BALQSOs already searched for microvariability could place
important constraints on the possible origin of such microvariations.

\section*{Acknowledgments}
The help rendered by the observer at the 1.04-m ARIES telescope,
Nainital, and technical staff at the 2.01-m HCT CREST is gratefully
acknowledged. We would like to thank the anonymous referee for 
constructive comments on the manuscript.

Funding for the SDSS and SDSS-II has been provided by the Alfred
P. Sloan Foundation, the Participating Institutions, the National
Science Foundation, the U.S. Department of Energy, the National
Aeronautics and Space Administration, the Japanese Monbukagakusho,
the Max Planck Society, and the Higher Education Funding Council for
England. The SDSS Web Site is http://www.sdss.org/.

\label{lastpage}
\end{document}